\documentclass[conference]{IEEEtran}
\usepackage{blindtext, graphicx}
\usepackage[cmex10]{amsmath}
\usepackage{bm}
\usepackage{amssymb,amsthm,mathrsfs,amsfonts,amsmath}
\usepackage{bbm}
\usepackage{caption}

\PassOptionsToPackage{obeyspaces}{url}
\PassOptionsToPackage{bookmarks=false}{hyperref}

\usepackage{scalerel,stackengine}
\stackMath
\newcommand\reallywidehat[1]{%
	\savestack{\tmpbox}{\stretchto{%
			\scaleto{%
				\scalerel*[\widthof{\ensuremath{#1}}]{\kern-0.6pt\bigwedge\kern-.6pt}%
				{\rule[-\textheight/2]{1ex}{\textheight*4/2}}
			}{\textheight}%
		}{0.4ex}}%
	\stackon[1pt]{#1}{\tmpbox}%
}

\ifCLASSINFOpdf
\else
\fi

\IEEEoverridecommandlockouts
\begin{document}
%
\title{\huge Modelling of Sound Events with Hidden Imbalances Based on Clustering and Separate Sub-Dictionary Learning }
\vspace{-15pt}
\author{
	\IEEEauthorblockN{\textit{Chaitanya Narisetty, Tatsuya Komatsu \textnormal{and} Reishi Kondo}}
	\IEEEauthorblockA{Data Science Research Laboratories\\
		NEC Corporation, Japan\\
		Email: c-narisetty@cp.jp.nec.com, t-komatsu@ew.jp.nec.com, kondoh@ct.jp.nec.com}
}


%


\maketitle

\begin{abstract}
	This paper proposes an effective modelling of sound event spectra with a hidden data-size-imbalance, for improved Acoustic Event Detection (AED).
	The proposed method models each event as an aggregated representation of a few latent factors, while conventional approaches try to find acoustic elements directly from the event spectra.
	In the method, all the latent factors across all events are assigned comparable importance and complexity to overcome the hidden imbalance of data-sizes in event spectra.
	To extract latent factors in each event, the proposed method employs clustering and performs non-negative matrix factorization to each latent factor, and learns its acoustic elements as a sub-dictionary.
	Separate sub-dictionary learning effectively models the acoustic elements with limited data-sizes and avoids over-fitting due to hidden imbalances in training data.
	For the task of polyphonic sound event detection from DCASE 2013 challenge, an AED based on the proposed modelling achieves a detection F-measure of $46.5\%$, a significant improvement of more than $19\%$ as compared to the existing state-of-the-art methods.

	\vspace{5pt}
	
\end{abstract}

\begin{IEEEkeywords}
	Data-Size-Imbalance, Acoustic Event Detection, Non-Negative Matrix Factorization, Dictionary Learning
\end{IEEEkeywords}

%
\IEEEpeerreviewmaketitle

\section{Introduction}
\label{sec:Intro}
The ubiquity of recorded audio in the present day paves way for development of smart applications that can be readily integrated with voice activated devices like Amazon Echo or Google Home. Being invariant to occlusions and brightness, audio is ideally suited for security applications likes Acoustic Event Detection (AED) and Scene Identification \cite{stowell2015detection}. This paper focuses on AED which aims to detect the occurrences of sound events in a given audio signal. The event detection performance is significantly dependent on the effectiveness of modelling audio signals gathered for training. Hence the performance of AED degrades with noise, polyphony and hidden data-size-imbalances in audio signals. Polyphony refers to the simultaneous existence of two or more sound events in an audio signal. These degradations demand the need for effective event models with multi-label event classifiers that are robust to the interference from background noises.

\vspace{4pt}
The most frequently used approaches for AED are based on Mel-Frequency Cepstral Coefficients (MFCCs) trained using Gaussian Mixture Models \cite{vuegen2013mfcc} and Non-Negative Matrix Factorization (NMF) \cite{Zhou2017}. The recent advancements in AED demonstrate the effectiveness of Convolutional Recurrent Neural Networks (CRNNs) for modelling large audio databases \cite{Adavanne2017}\cite{Lim2017}. The capability of CRNNs to model complex and non-linear dependencies in the audio signals is offset by its challenges of being computationally intensive and often prone to severe over-fitting \cite{hinton2012deep}. These challenges motivate the use of matrix factorization methods to extract linear dependencies in audio signals. The additional constraint of having non-negative factors is to model a given audio as an additive representation of different sound sources. The entire set of linear dependencies (dictionary) among all sound events can be extracted at once by performing NMF on the spectrogram of an entire audio database, consisting of several sound events \cite{cotton2011spectral}. Hence the performance of NMF based AED methods heavily rely on the ability of their dictionaries to represent sound events. In situations where the data-size of a few events is much higher than the others, dictionary learned by performing NMF of the entire database overlooks the events with limited data-sizes. One of the methods to model such an imbalance is to normalize the spectra of each event with their respective data-sizes \cite{guillamet2001weighted}. It is also possible to estimate the overall dictionary as a composition of event-wise dictionaries to overcome the imbalance among sound events \cite{gemmeke2013exemplar}.

\vspace{4pt}
A similar but hidden imbalance among the exemplars of an individual sound event was observed in NEC's field trials \cite{komatsu2017acoustic}. For the purpose of illustration, consider a \textit{Piano} sound event consisting of $10$ exemplars for 3 notes $C4$-$G4$-$E4$ played simultaneously for $1$ second and only $1$ exemplar for a $C4$ note played for $0.5$ seconds. \textit{Piano} event dictionary learned using NMF fails to model the under-represented $C4$ note. If prior information detailing the imbalance of exemplars is hidden, the dictionary learned for such an event fails to represent the exemplars with limited data-sizes.

\vspace{4pt}
We propose an effective dictionary learning of sound events to overcome such data-size-imbalances by assuming an event to be explained from a few latent factors.
This assumption motivates an effective strategy for dictionary generation where each event is clustered into latent factors namely sub-events. Each sub-event is an aggregation of multiple acoustic elements (a structural element which can be represented by a single basis). Clustering brings forth the underlying data-size-imbalance hidden in each event. Sub-dictionaries learned from each sub-event improve the modelling of under-represented acoustic elements. The overall set of sub-dictionaries over all sound events effectively models the entire audio database, and when paired with support vector machines, improves on the drawback of existing NMF based AED methods.

\vspace{4pt}
The remainder of this paper is organized as follows. In Section~\ref{sec:NMF} we briefly discuss a conventional NMF based dictionary learning. Section~\ref{sec:PropMeth} describes an AED method based on the proposed dictionary learning and SVM classifier training. Performance evaluations of the proposed method for the above \textit{Piano} illustration and the Polyphonic Sound Event Detection task from Detection and Classification of Acoustic Scenes and Events (DCASE) 2013 are detailed in Section~\ref{sec:SimsResults}. This discussion ends with a few concluding remarks in Section~\ref{sec:Conc}.

\section{Non-Negative Matrix Factorization}
\label{sec:NMF}
Before detailing the proposed dictionary learning method, we briefly explain the concept of NMF as it is predominantly used for modelling of sound events. 
NMF is a set of matrix decomposition techniques which approximately factorize a given positive matrix $V$ of size $p\times q$ into positive low-rank matrices $W$ and $H$ of sizes $p\times r$ and $r\times q$ respectively, where $r < \min(p, q)$. It is being assumed that $V$ consists of $q$ features each of which is a $p$-dimensional vector. In the context of an audio signal obtained from an additive mixture of sound sources, the audio spectrogram $V$ can be factorized into a set of few basis vectors $W$ and activation vectors $H$. NMF is formulated as,
\begin{align}
V \approx WH \text{ s.t. } V \succeq 0, W\succeq 0, H \succeq 0.
\end{align}
This approximate factorization is guided by a cost function minimization. For our present work, Kullback-Leibler (KL) Divergence is taken to be the cost function. Iterative update rules used to estimate $W$ and $H$ for NMF optimized using KL-Divergence are put forth by Lee and Seung \cite{lee2001algorithms}.

\vspace{5pt}
\subsubsection*{Problem with NMF dictionaries} $W$ represents a dictionary which consists of fundamental vectors necessary to reconstruct $V$. Columns of $H$ represent a new feature space and indicate the parts of dictionary $W$ that are being used to represent a given spectra \cite{joder2012exploring}. A natural application of such dictionary dependent features $H$ is to identi	fy activations extracted from a given test spectra, that are similar to activations learned from training spectra.
Without a proper dictionary, the activations necessary to represent a spectrum cannot be obtained. From the spectrogram of entire \textit{Piano} event described earlier, a dictionary learned using NMF aims to represent most parts of the spectra and ends up over-fitting the $C4$-$G4$-$E4$ sound. Thus there are no basis vectors capable of modelling the $C4$ sound. To overcome this inability of conventional NMF to represent hidden acoustic elements with limited data-sizes, we propose an improved dictionary learning method.

\section{AED with Proposed Dictionary Learning}
\label{sec:PropMeth}
\subsection{Dictionary Learning}
\begin{figure}[t]
	\centering
	\includegraphics[clip, trim=0.5cm 4cm 0.5cm 0cm, width=\linewidth]{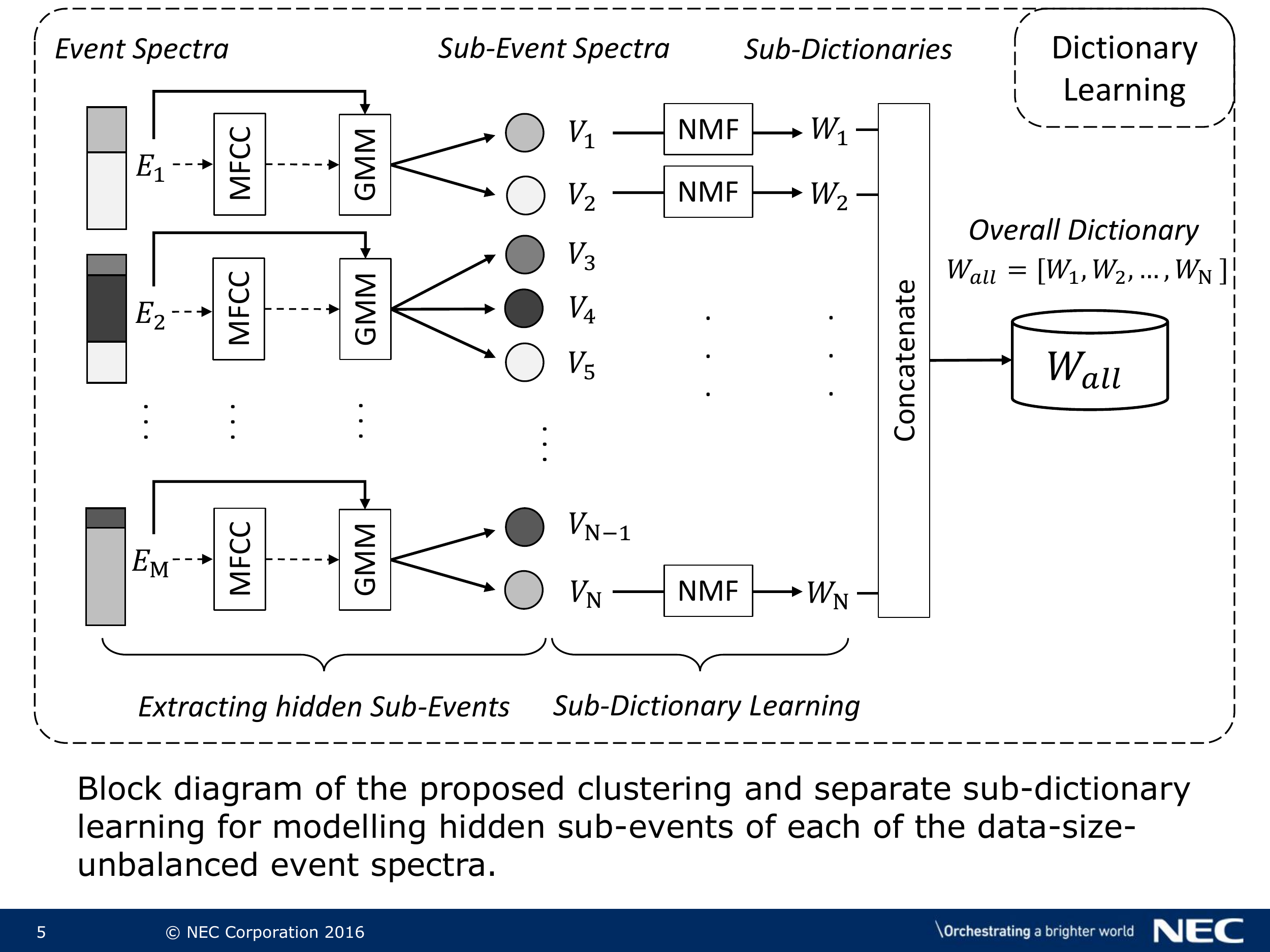}
	\caption{Block diagram of the proposed clustering and separate sub-dictionary learning for modelling the hidden imbalance of data-sizes in each event spectra.}
	\label{fig:BlockDiagram}
	\vspace{-10pt}
\end{figure}

Block diagram for the proposed method consists of two important parts: clustering event spectra and separate sub-dictionary learning as shown in Fig.~\ref{fig:BlockDiagram}. Consider an audio database of $M$ sound events with each event's respective spectrogram denoted by $E_m$, $1\leq m \leq M$. MFCC coefficients for each spectrum in $E_m$ are extracted and trained using a Gaussian mixture model (GMM). Each of the event spectra $E_m$ is clustered into a set of sub-event spectra using the trained GMM. Let the set $\{V_n\}, 1\leq n \leq N$ represent the entire set of sub-event spectra, where $N$ is the total number of sub-events clustered across all events. Separately, a sub-dictionary $W_n$ is learned using NMF from their respective sub-event spectra $V_n$. All sub-dictionaries are concatenated to output an overall dictionary $W_{all}$ for the entire audio database i.e.
\begin{align}
W_{all} = [ W_1, W_2, \dots W_N ].
\end{align}

The motivation behind clustering event spectra is to extract the hidden data-size-imbalances among its acoustic elements. In the \textit{Piano} event illustrated earlier, the unbalanced acoustic elements are the well-represented spectra of $C4$-$G4$-$E4$ and under-represented spectra of $C4$. As it is not always possible to construct event databases which take such imbalance into consideration, mixture models such as GMM are capable of modelling both the under-represented and well-represented parts of an event \cite{reynolds1995robust}. The extracted MFCCs reduce the dimensionality of original spectrum and improve the stability of mixture modelling. Separately learned sub-dictionaries make $W_{all}$ capable of representing the acoustic elements with limited data-sizes, which are overlooked by the conventional event-wise NMF dictionaries. Hence an AED based on the proposed dictionary learning effectively models sound events, thereby improving the overall detection performance.
%

\subsection{Classifier Training and Event Detection}
\begin{figure}
	\centering
	\includegraphics[clip, trim=1.5cm 7cm 2cm 1.5cm, width=\linewidth]{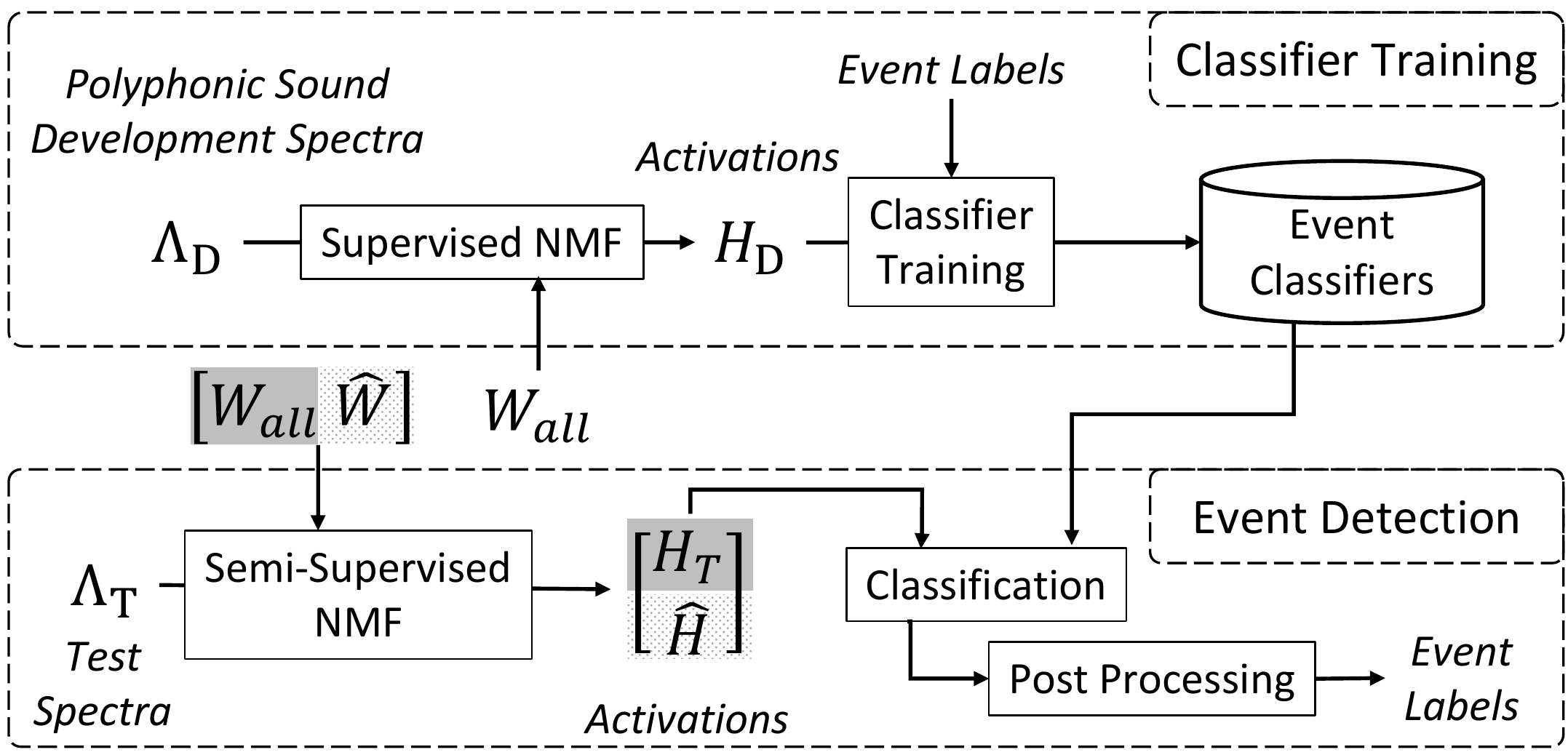}
	\caption{Block diagram for training event classifiers from the labelled spectra of polyphonic sound events and using these classifiers to detect sound event labels from test spectra.}
	\label{fig:TrainTest}
	\vspace{-10pt}
\end{figure}
The above dictionary learning is an integral part of the overall acoustic event modelling and detection as shown in Fig.~\ref{fig:TrainTest}. Let  $\Lambda_D$ be the development spectra used for training classifiers. In this work, $\Lambda_D$ is the spectrogram of polyphonic audio signals obtained from a mixture of different sound events. Supervised NMF is performed over the entire development spectra $\Lambda_D$ by using a fixed basis matrix $W_{all}$ to estimate the activations matrix $H_D$. A conventional approach to train event classifiers involves the use of this activations matrix $H_D$. The columns of $H_D$, along with their respective event labels are used for classifier training to generate event classifiers.

\vspace{4pt}
For detecting sound events in any given test spectra $\Lambda_T$, a semi-supervised NMF with an extended basis matrix $[W_{all}, \reallywidehat{W}]$ is employed \cite{weninger2012supervised}. As the test spectra often contains an added noise, a small noise dictionary $\reallywidehat{W}$ is appended to the columns of $W_{all}$. During this semi-supervised NMF, $W_{all}$ remains fixed and only $\reallywidehat{W}$ is estimated. The extended portion of the basis matrix $\reallywidehat{W}$ is estimated using NMF to model the additional noise elements. This increment of the overall basis matrix increases the dimensionality of the estimated activations matrix. This is formulated as,
\begin{align}
\Lambda_T \approx \begin{bmatrix}W_{all} & \reallywidehat{W}\end{bmatrix} \begin{bmatrix}H_T \vspace{5pt}\\ \reallywidehat{H}\end{bmatrix} \text{ s.t. } \Lambda_T, \reallywidehat{W}, H_T, \reallywidehat{H} \succeq 0.
\end{align}
The part of activations matrix corresponding to $W_{all}$ is $H_T$ and that of $\reallywidehat{W}$ is $\reallywidehat{H}$. Only the event activation matrix $H_T$ is used for event detection, while $\reallywidehat{H}$ is ignored. The trained event classifiers classify each column of $H_T$ to output a binary eventroll that indicates which event(s) take place in each column of $H_T$. The extracted binary eventroll is later post-processed to output the final event labels.

\vspace{4pt}
Many types of classifiers are used in literature to train these activations. Gemmeke et al. \cite{gemmeke2013exemplar} uses a Hidden Markov Model (HMM) with linear time warping. Thresholding based classifiers to ascertain the existence of a particular event have been used in \cite{wang2017non}. However, Support Vector Machines (SVMs) with linear kernels have shown promising results for training event classifiers \cite{komatsu2016acoustic, weninger2013discriminative, zilu2009facial} and will be used in this paper. The activations corresponding to a particular event are used to train a binary linear-SVM classifier against all the activations of the remaining events. 


\section{Simulations and Results}
\label{sec:SimsResults}
\subsection{Dictionary Learning for the \textit{Piano} Event}
\begin{figure}
	\centering
	\includegraphics[clip, trim=0cm 0.5cm 0cm 0cm, width=\linewidth]{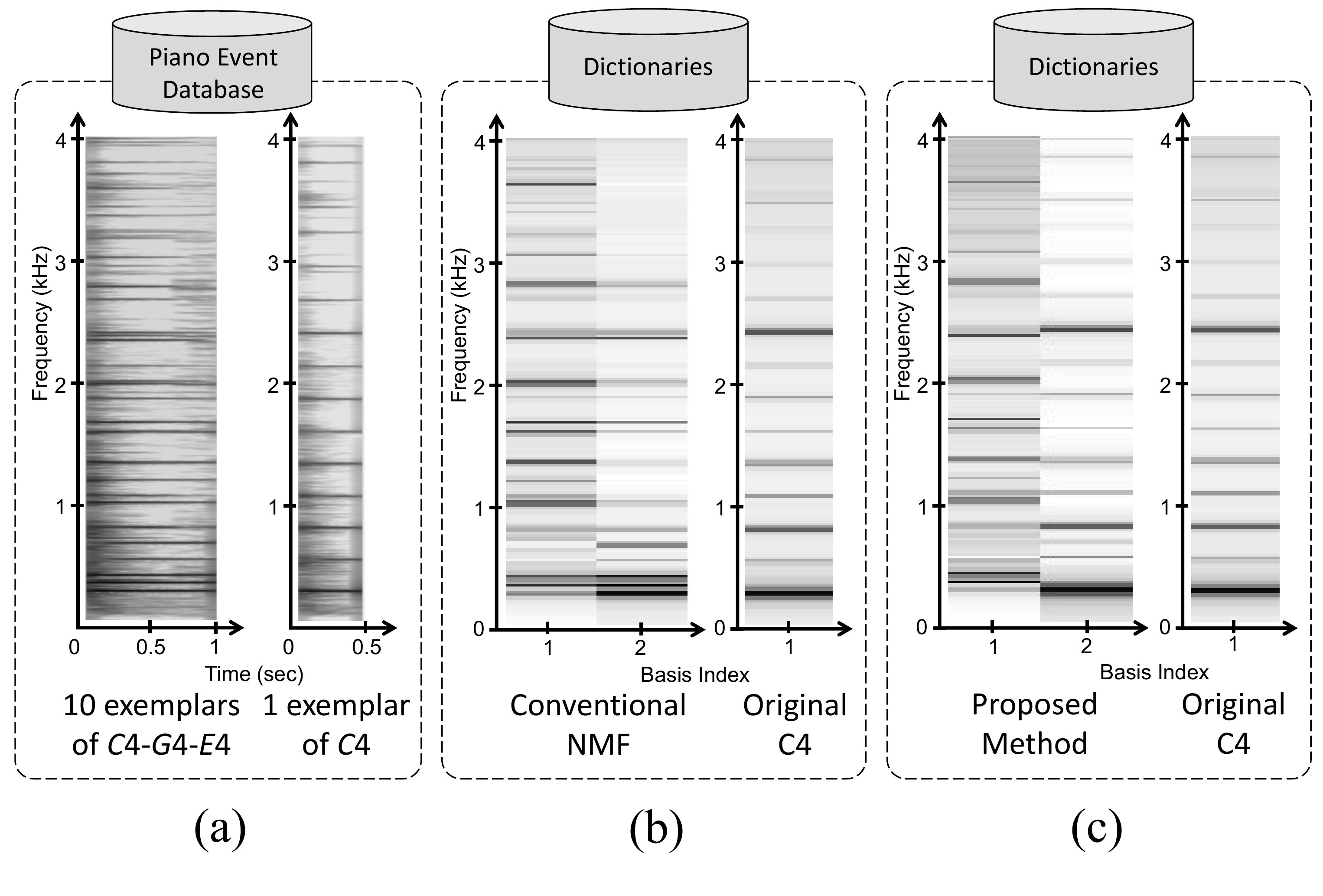}
	\caption{(a) \textit{Piano} sound event with $20:1$ data-size-imbalance of spectra and its dictionaries extracted using (b) Conventional NMF which can collectively represent only the  $C4$-$G4$-$E4$ and (c) Proposed Method which can represent both $G4$-$E4$ and $C4$.}
	\label{fig:ConvPropDictComp}
	\vspace{-15pt}
\end{figure}

As detailed in the previous sections, consider the \textit{Piano} event with a data-size-imbalance among its $C4$-$G4$-$E4$ and $C4$ exemplars as shown in Fig.~\ref{fig:ConvPropDictComp}(a). Ideally, the dictionary for this event should contain two basis vectors which are either $\{C4$-$G4$-$E4$, $C4\}$ or $\{G4$-$E4$, $C4\}$. The sampling rate of the exemplars is $8$kHz. The spectrogram for this event is estimated from $40$ms time frames of the signal with a $10$ms frame shift.

\vspace{4pt}
When a dictionary is learned from this spectrogram using conventional NMF, the resulting two basis vectors are low and high frequency representations of the $C4$-$G4$-$E4$ sound and are unable to represent the $C4$ sound as shown in Fig.~\ref{fig:ConvPropDictComp}(b). A single basis vector obtained from the $C4$ exemplar using NMF is shown on the side for visual comparison. Alternatively using the proposed method, we first extract $25$ MFCCs excluding the $0^{th}$ coefficient from each spectrum. A GMM with two components models the MFCCs and clusters the entire event into two sub-event spectra. Then sub-dictionaries with two basis vectors are learned from each sub-event. The resulting $W_{all}$ contains a total of four basis vectors. There are only two distinct acoustic elements ($C4$, $C4$-$G4$-$E4$) in the overall event, so the dimensionality of $W_{all}$ can be reduced to $2$ by identifying the two basis vectors with minimum correlation between them. These are shown in Fig.~\ref{fig:ConvPropDictComp}(c) and it can be seen that the second basis vector closely resembles the single basis vector obtained from $C4$ exemplar. Also, both basis vectors put together closely represent the $C4$-$G4$-$E4$ sound. This demonstrates the ability of the proposed method in learning dictionaries that represent the entire sound event.

\begin{figure}
	\centering
	\includegraphics[clip, trim=0cm 0cm 0cm 0cm, width=\linewidth]{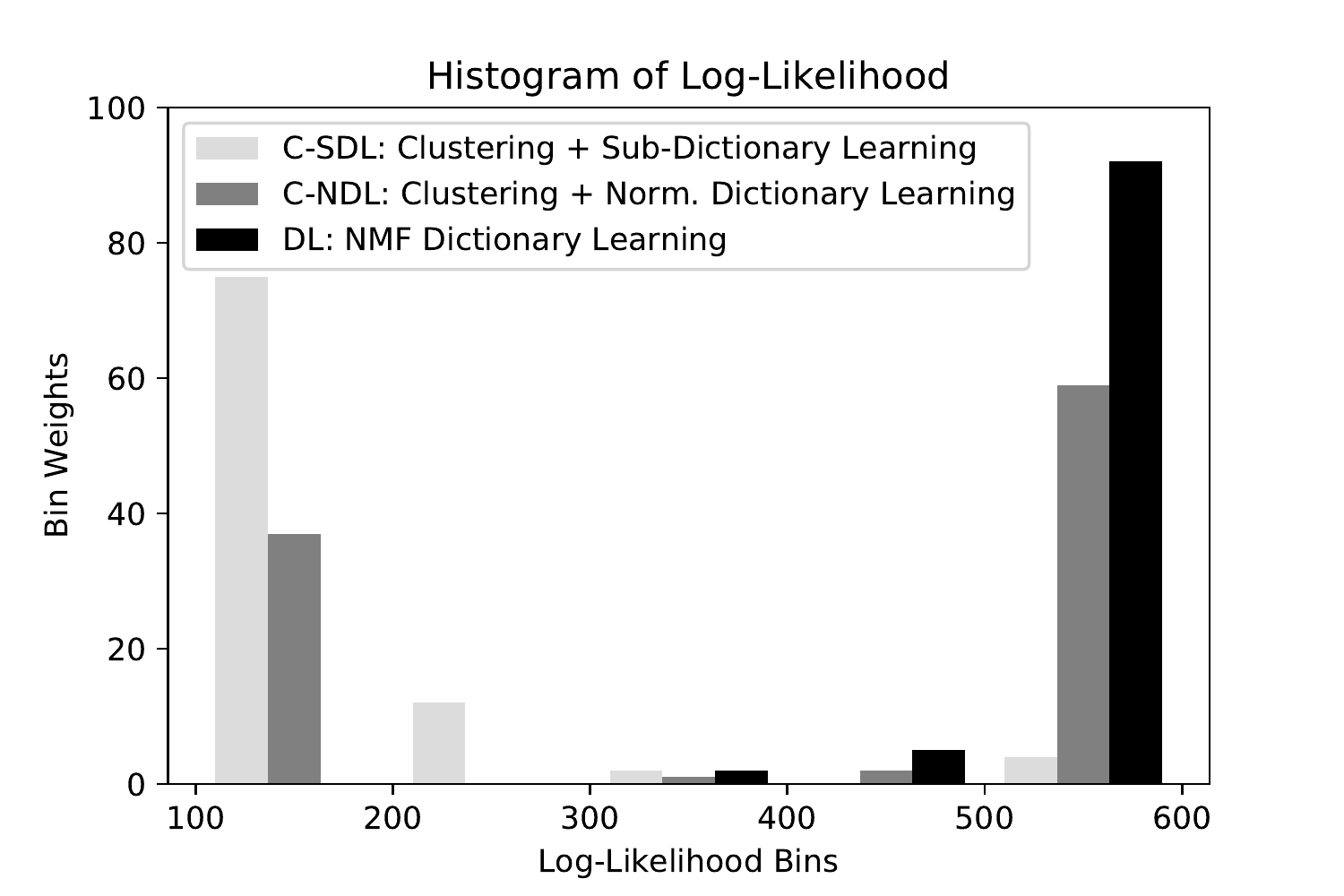}\vspace{2pt}
	\caption{Histogram of Log-Likelihoods from the reconstruction of the $C4$ exemplar from dictionaries extracting using C-SDL, C-NDL and DL methods. The first and last histogram bins indicate accurate and failed reconstructions respectively.}
	\label{fig:histErrors}
	\vspace{-5pt}
\end{figure}

\vspace{4pt}
For realistic sound events, GMM based clustering of event spectra is dependent on the initialization parameters, and therefore not reliable in modelling unbalanced acoustic elements accurately. If clustering of event spectra can separate the spectra of $C4$-$G4$-$E4$ and $C4$ accurately, normalizing each of the clustered sub-event spectra by their data-size should be sufficient for NMF to extract an effective dictionary from the entire normalized event spectra \cite{guillamet2001weighted}. Hence we additionally compare the proposed method with clustering and normalized sub-dictionary learning. We evaluate the reconstruction error (log-likelihood) of the $C4$ exemplar at $100$ random initializations using the proposed dictionary learning, clustering and normalized dictionary learning and the conventional NMF dictionary learning. A histogram of the $100$ log-likelihoods for each method is shown in Fig.~\ref{fig:histErrors}. Among the $5$ histogram bins, the first bin indicates an accurate reconstruction and the last bin indicates a failed reconstruction. It is evident that the conventional NMF fails to represent the $C4$ exemplar at almost all initializations. The clustering and normalized dictionary learning succeeds to reconstruct at less than $40\%$ of all initializations, while the proposed dictionary learning succeeds at more than $75\%$. This demonstrates the robustness of the proposed method to uncertainties in clustering effective sub-event spectra.

\subsection{Polyphonic Sound Event Detection}
\subsubsection*{Training and Testing}
In this paper, we evaluate the proposed method for polyphonic sound event detection task from the IEEE challenge for DCASE 2013 \cite{giannoulis2013detection}. The dictionaries for each sound event are learned from the isolated training database consisting of $M=16$ sound events. Event spectra are estimated from $40$ms frame sizes with a $10$ms frame shift. Each event spectra is first clustered into $2$ sub-event spectra by extracting $23$ MFCCs excluding the $0^{th}$ coefficient and then modelling using a $2$ component GMM. Then sub-dictionaries with $3$ basis vectors are learned from each of the clustered sub-event spectra. The size of overall dictionary $W_{all}$ is $N=96$.

\vspace{4pt}
Spectra of OS (Office Synthetic) development database consisting of $9$ polyphonic synthetic recordings ($90$ seconds each) is used for classifier training. Note that the dimensionality of MFCCs, GMM and sub-dictionaries are optimized for the OS development database. Due to a large number of sound events and polyphonic nature of the development database, class weight for event activations to be trained using binary linear-SVM classifiers is increased to $3$ times. Similar to the training phase, test spectra is obtained from OS test database which contains $12$ recordings ($120$ seconds each) with different levels of polyphony and background noises. A second database OS-IRCCYN \cite{osirccyn} is also used for evaluating the proposed method. Both databases share the same ground truths, however the latter is obtained from audio recorded at IRCCYN, France \cite{benetos2017polyphonic}. Size of the noise dictionary $\reallywidehat{W}$ is $1$. For post-processing, the generated binary eventroll is filtered using a median filter of length $3$. Gaps less than $250$ms in the eventroll are filled and the final event labels with duration more than $200$ms are considered to be valid.

\vspace{4pt}
\subsubsection*{Evaluations and Comparisons}
$\mathcal{F}$-measure based metrics are used evaluation in this paper. We consider three different evaluation metrics: a $10$ms frame-based $\mathcal{F}$-measure ($\mathcal{F}_{fb}$) from DCASE 2013, $100$ms segment-based ($\mathcal{F}_{sb}$) and class-wise segment-based $\mathcal{F}$-measures ($\mathcal{F}_{cwsb}$) from DCASE 2016. Frame-based metrics are evaluated for each recording and their average is noted, while segment-based metrics are evaluated over the entire database.

\begin{figure}
	\centering
	\includegraphics[clip, trim=3cm 6.5cm 4.5cm 0.5cm, width=\linewidth]{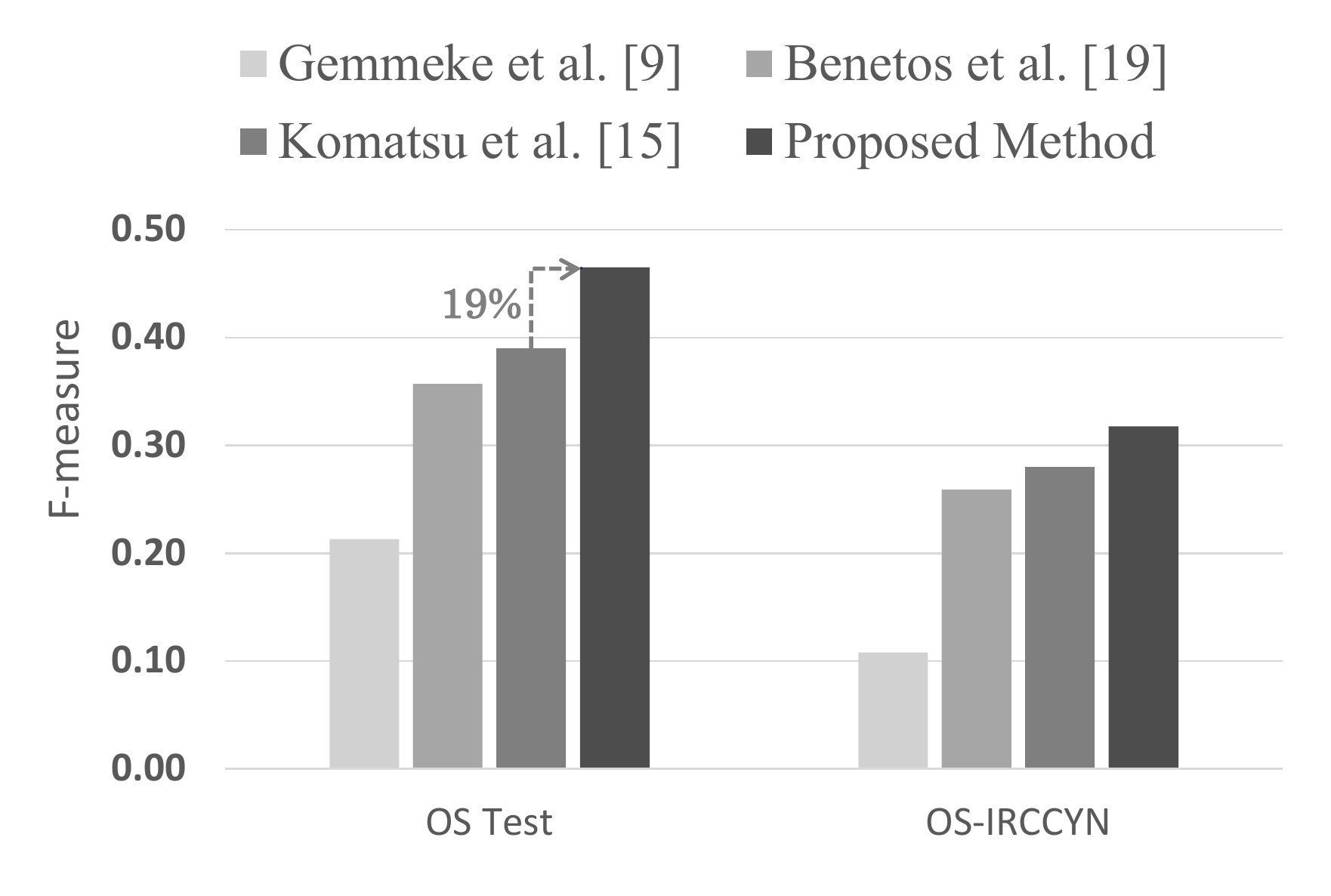}
	\caption{Comparison of Frame-based $\mathcal{F}$-measures for the polyphonic event detection based on the proposed modelling of sound events for OS Test and OS-IRCCYN Databases with current state-of-the-art methods.}
	\label{fig:MethodsComp}
	\vspace{-5pt}
\end{figure}

\vspace{4pt}
To evaluate the proposed method, we compare it with three most relevant NMF based AED methods. First method proposed by Gemmeke et al. uses an event-wise NMF for dictionary learning and extracts event-likelihoods using Hidden Markov Models (HMM) with linear time warping \cite{gemmeke2013exemplar}. Second is the probabilistic latent component analysis (PLCA) with integrated linear dynamical systems (LDS) proposed by Benetos et al. \cite{benetos2017polyphonic} for polyphonic sound event tracking. The third is the sparsely activated mixture of local dictionaries (MLD) based dictionary learning proposed by Komatsu et al. and trained using SVM classifiers \cite{komatsu2016acoustic}. For this comparison, the three NMF based AED methods are evaluated for both OS test and OS-IRCCYN databases, and their $\mathcal{F}_{fb}$ are shown in Fig.~\ref{fig:MethodsComp}. An AED based on the proposed dictionary learning achieves a frame-based $\mathcal{F}$-measure of $46.5\%$ for the OS test database and shows a significant improvement of $19.2\%$ over the next best state-of-the-art AED method. The proposed dictionary learning has improved event modelling thereby outperforming the existing NMF based state-of-the-art methods. To the best of our knowledge, the reported $\mathcal{F}_{fb}$ of $46.5\%$ for OS test database using the proposed method is highest in existing literature.

\begin{figure}
	\centering
	\includegraphics[clip, trim=3.7cm 6.5cm 4.2cm 1cm, width=\linewidth]{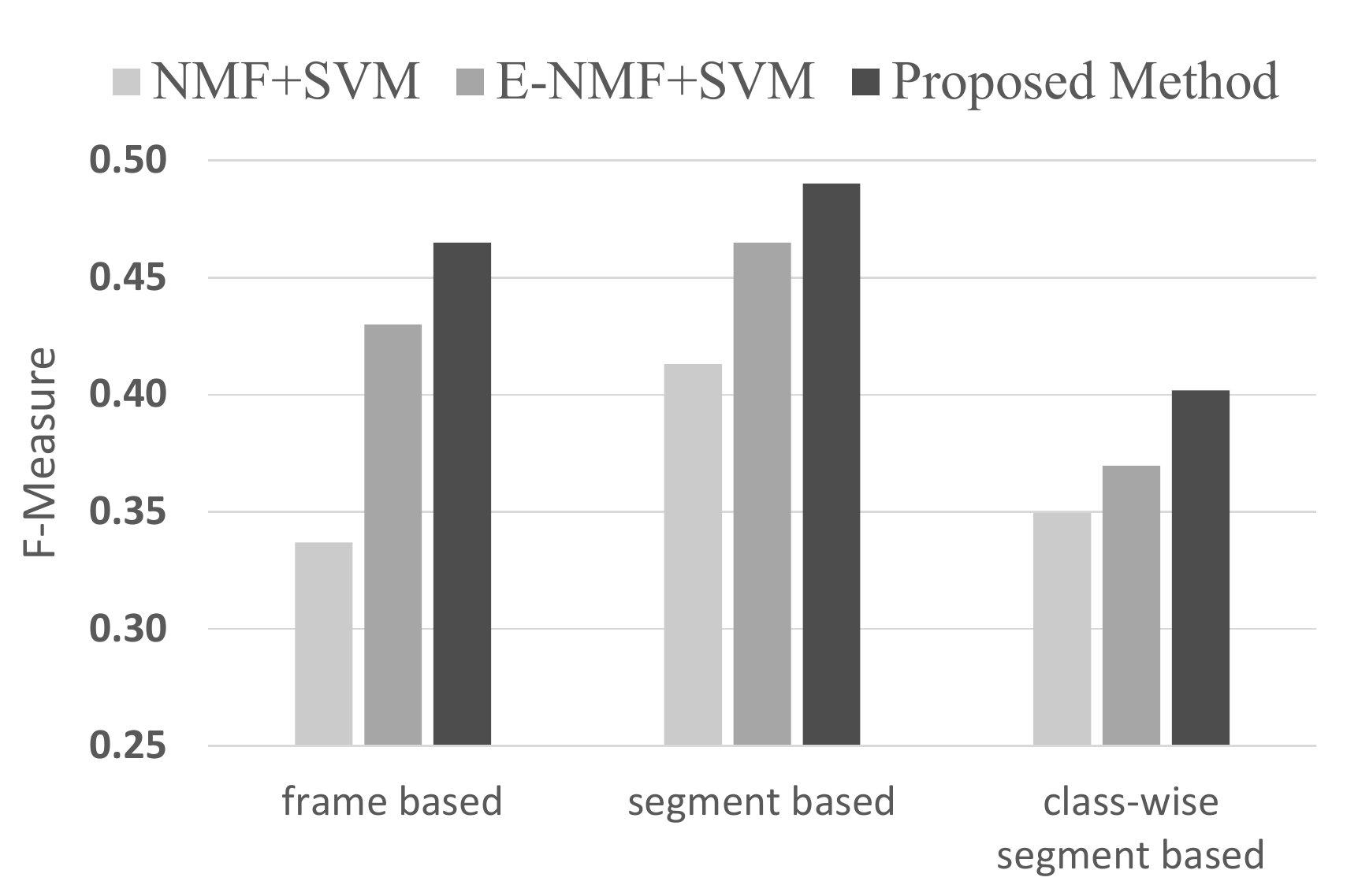}
	\caption{Significance of the proposed modelling of sound events across various $\mathcal{F}$-measure metrics for the polyphonic DCASE 2013 OS Test Database.}
	\label{fig:FmsComp}
	\vspace{-10pt}
\end{figure}

\vspace{4pt}
Further evaluating the significance of proposed dictionary learning, we compare it with two conventional NMF-SVM based AED methods. First method learns the overall dictionary with $35$ basis vectors by doing NMF to the entire audio spectra. The second method performs an event-wise NMF and groups together all the event dictionaries. Size of these event dictionaries are tested from $1$ to $6$, with $3$ being the best. The two conventional NMF based AED methods are denoted as NMF+SVM and E-NMF+SVM respectively. A binary linear-SVM classifier is trained in all the three methods. $\mathcal{F}$-measures for the three methods are shown in Fig.~\ref{fig:FmsComp}. The proposed method achieves the highest $\mathcal{F}_{fb}, \mathcal{F}_{sb}, \mathcal{F}_{cwsb}$ of $46.5\%, 49\%$ and $40.2\%$ respectively. This comparison shows the significance of learning sub-dictionaries separately.

\section{Conclusions}
\label{sec:Conc}
In this work, we propose an effective dictionary learning of sound events with hidden data-size-imbalances, for the task of acoustic event detection. The imbalance among data-sizes of acoustic elements present in the spectra of each sound event are first estimated by clustering and the sub-dictionaries from each cluster spectra are learned separately. An illustrative example shows that the overall set of sub-dictionaries is better able to model the hidden imbalances as compared to conventional NMF based dictionary learning methods. We further demonstrate the superiority of an AED method based on the proposed dictionary learning, over three existing state-of-the-art AED methods. Rigorous mathematical work integrating the clustering and sub-dictionary learning into the NMF formulation is left for future work.

\bibliographystyle{IEEEtran}
\bibliography{mybib}{}

\begin{thebibliography}{10}
\providecommand{\url}[1]{#1}
\csname url@samestyle\endcsname
\providecommand{\newblock}{\relax}
\providecommand{\bibinfo}[2]{#2}
\providecommand{\BIBentrySTDinterwordspacing}{\spaceskip=0pt\relax}
\providecommand{\BIBentryALTinterwordstretchfactor}{4}
\providecommand{\BIBentryALTinterwordspacing}{\spaceskip=\fontdimen2\font plus
\BIBentryALTinterwordstretchfactor\fontdimen3\font minus
  \fontdimen4\font\relax}
\providecommand{\BIBforeignlanguage}[2]{{%
\expandafter\ifx\csname l@#1\endcsname\relax
\typeout{** WARNING: IEEEtran.bst: No hyphenation pattern has been}%
\typeout{** loaded for the language `#1'. Using the pattern for}%
\typeout{** the default language instead.}%
\else
\language=\csname l@#1\endcsname
\fi
#2}}
\providecommand{\BIBdecl}{\relax}
\BIBdecl

\bibitem{stowell2015detection}
D.~Stowell, D.~Giannoulis, E.~Benetos, M.~Lagrange, and M.~D. Plumbley,
  ``Detection and classification of acoustic scenes and events,'' in \emph{IEEE
  Transactions on Multimedia}, vol.~17, no.~10, 2015, pp. 1733--1746.

\bibitem{vuegen2013mfcc}
L.~Vuegen, B.~Broeck, P.~Karsmakers, J.~Gemmeke, B.~Vanrumste, and H.~Hamme,
  ``An mfcc-gmm approach for event detection and classification,'' in
  \emph{{IEEE} Workshop on Applications of Signal Processing to Audio and
  Acoustics (WASPAA)}, 2013, pp. 1--3.

\bibitem{Zhou2017}
Q.~Zhou and Z.~Feng, ``Robust sound event detection through noise estimation
  and source separation using {NMF},'' DCASE2017 Challenge, Tech. Rep.,
  September 2017.

\bibitem{Adavanne2017}
S.~Adavanne and T.~Virtanen, ``A report on sound event detection with different
  binaural features,'' DCASE2017 Challenge, Tech. Rep., September 2017.

\bibitem{Lim2017}
H.~Lim, J.~Park, and Y.~Han, ``Rare sound event detection using {1D}
  convolutional recurrent neural networks,'' DCASE2017 Challenge, Tech. Rep.,
  September 2017.

\bibitem{hinton2012deep}
G.~Hinton, L.~Deng, D.~Yu, G.~E. Dahl, A.~R. Mohamed, N.~Jaitly, A.~Senior,
  V.~Vanhoucke, P.~Nguyen, T.~N. Sainath \emph{et~al.}, ``Deep neural networks
  for acoustic modeling in speech recognition: The shared views of four
  research groups,'' in \emph{{IEEE} Signal Processing Magazine}, vol.~29,
  no.~6, 2012, pp. 82--97.

\bibitem{cotton2011spectral}
C.~V. Cotton and D.~P. Ellis, ``Spectral vs. spectro-temporal features for
  acoustic event detection,'' in \emph{{IEEE} Workshop on Applications of
  Signal Processing to Audio and Acoustics (WASPAA)}, 2011, pp. 69--72.

\bibitem{guillamet2001weighted}
D.~Guillamet, M.~Bressan, and J.~Vitria, ``A weighted non-negative matrix
  factorization for local representations,'' in \emph{{IEEE} Proceedings of
  Computer Society Conference on Computer Vision and Pattern Recognition
  (CVPR)}, vol.~1, 2001, pp. I--I.

\bibitem{gemmeke2013exemplar}
J.~F. Gemmeke, L.~Vuegen, P.~Karsmakers, B.~Vanrumste \emph{et~al.}, ``An
  exemplar-based nmf approach to audio event detection,'' in \emph{{IEEE}
  Workshop on Applications of Signal Processing to Audio and Acoustics
  (WASPAA)}, 2013, pp. 1--4.

\bibitem{komatsu2017acoustic}
T.~Komatsu, M.~Tani, T.~Toizumi, N.~Chaitanya, M.~Kato, Y.~Arai, O.~Hoshuyama,
  Y.~Senda, and R.~Kondo, ``An acoustic monitoring system and its field
  trials,'' in \emph{Asia-Pacific Signal and Information Processing Association
  Annual Summit and Conference (APSIPA ASC)}.\hskip 1em plus 0.5em minus
  0.4em\relax IEEE, 2017, pp. 1341--1346.

\bibitem{lee2001algorithms}
D.~D. Lee and H.~S. Seung, ``Algorithms for non-negative matrix
  factorization,'' in \emph{Advances in neural information processing systems
  (NIPS)}, 2001, pp. 556--562.

\bibitem{joder2012exploring}
C.~Joder and B.~Schuller, ``Exploring nonnegative matrix factorization for
  audio classification: Application to speaker recognition,'' in \emph{{VDE}
  Proceedings of Speech Communication; 10. ITG Symposium}, 2012, pp. 1--4.

\bibitem{reynolds1995robust}
D.~A. Reynolds and R.~C. Rose, ``Robust text-independent speaker identification
  using gaussian mixture speaker models,'' in \emph{{IEEE} transactions on
  speech and audio processing}, vol.~3, no.~1, 1995, pp. 72--83.

\bibitem{weninger2012supervised}
F.~Weninger, J.~Feliu, and B.~Schuller, ``Supervised and semi-supervised
  suppression of background music in monaural speech recordings,'' in
  \emph{{IEEE} International Conference on Acoustics, Speech and Signal
  Processing (ICASSP)}, 2012, pp. 61--64.

\bibitem{wang2017non}
S.~Wang and J.~Ortiz, ``Non-negative matrix factorization of signals with
  overlapping events for event detection applications,'' in \emph{{IEEE}
  International Conference on Acoustics, Speech and Signal Processing
  (ICASSP)}, 2017, pp. 5960--5964.

\bibitem{komatsu2016acoustic}
T.~Komatsu, T.~Toizumi, R.~Kondo, and Y.~Senda, ``Acoustic event detection
  method using semi-supervised non-negative matrix factorization with a mixture
  of local dictionaries,'' in \emph{Proceedings of the Detection and
  Classification of Acoustic Scenes and Events 2016 Workshop (DCASE)}, 2016,
  pp. 45--49.

\bibitem{weninger2013discriminative}
F.~Weninger, C.~Kirst, B.~Schuller, and H.-J. Bungartz, ``A discriminative
  approach to polyphonic piano note transcription using supervised non-negative
  matrix factorization,'' in \emph{{IEEE} International Conference on
  Acoustics, Speech and Signal Processing (ICASSP)}, 2013, pp. 6--10.

\bibitem{zilu2009facial}
Y.~Zilu and Z.~Guoyi, ``Facial expression recognition based on nmf and svm,''
  in \emph{{IEEE} International Forum on Information Technology and
  Applications (IFITA)}, vol.~3, 2009, pp. 612--615.

\bibitem{giannoulis2013detection}
D.~Giannoulis, E.~Benetos, D.~Stowell, M.~Rossignol, M.~Lagrange, and M.~D.
  Plumbley, ``Detection and classification of acoustic scenes and events: An
  ieee aasp challenge,'' in \emph{{IEEE} Workshop on Applications of Signal
  Processing to Audio and Acoustics (WASPAA)}, 2013, pp. 1--4.

\bibitem{osirccyn}
E.~Benetos, ``{OS-IRCCYN datasets for Sound Event Detection},''
  \url{https://archive.org/details/OS-IRCCYN/}, [Online; accessed
  14-June-2018].

\bibitem{benetos2017polyphonic}
E.~Benetos, G.~Lafay, M.~Lagrange, M.~D. Plumbley, E.~Benetos, G.~Lafay,
  M.~Lagrange, and M.~D. Plumbley, ``Polyphonic sound event tracking using
  linear dynamical systems,'' in \emph{{IEEE/ACM} Transactions on Audio, Speech
  and Language Processing (TASLP)}, vol.~25, no.~6, 2017, pp. 1266--1277.

\end{thebibliography}
	
\end{document}